\documentclass[12pt]{iopart}

\usepackage{graphicx}
\usepackage{dcolumn}
\usepackage{bm}
\begin{document}

\title[How to calculate the fractal dimension of a complex network]{How to calculate the fractal dimension of a complex network: the box covering algorithm}

\author{Chaoming Song$^1$, Lazaros K. Gallos$^1$, Shlomo Havlin$^2$, Hern\'an A. Makse$^1$}

\address{$^1$ Levich Institute and Physics Department, City
College of New York,
New York, NY 10031, USA
$^2$ Minerva Center and Department of Physics, Bar-Ilan University,
52900 Ramat-Gan, Israel}

\date{\today}

\begin{abstract}
Covering a network with the minimum possible number of boxes can reveal interesting features
for the network structure, especially in terms of self-similar or fractal characteristics.
Considerable attention has been recently devoted to this problem, with the finding
that many real networks are self-similar fractals.
Here we present, compare and study in detail a number of algorithms that we
have used in previous papers towards this goal.
We show that this problem can be mapped to the well-known graph coloring problem
and then we simply can apply well-established algorithms. This seems to be the most
efficient method, but we also present two other algorithms based on burning
which provide a number of other benefits.
We argue that the presented algorithms provide a solution close to optimal and that another
algorithm that can significantly improve this result in an efficient way does not exist.
We offer to anyone that finds such a method to cover his/her
expenses for a 1-week trip to our lab in New York (details in http://jamlab.org).
\end{abstract}

\maketitle

\section{Introduction}

Complex networks are important since they describe efficiently many social,
biological and communication systems \cite{AB,DM,Newman,Pastor,BLMCH}. There exist many types of networks and
characterizing their topology is very important for a wide
range of static and dynamic properties. Recently \cite{shm,shm2}, we applied
a box covering algorithm which enabled us to demonstrate the existence
of self-similarity in many real networks. The fractal and self-similarity properties
of complex networks were subsequently studied extensively in a variety
of systems \cite{Yook,Palla,Zhao,Goh,Xu,Barriere,Carmi,Moreira,Estrada,Guida}.
In this paper we provide
a detailed study of the algorithms used to calculate
quantities characterizing the topology of such networks, such as the
fractal dimension $d_B$.
We study and compare several possible box covering algorithms,
by applying them to a number of model and real-world networks and we relate the
box covering optimization to the well-known vertex coloring algorithm \cite{JensenBook}. We
also suggest a new definition for the box size $\ell_B$, which seems to
yield more accurate values for the fractal dimension $d_B$ of a complex
network.

We show that the optimal network covering can be directly mapped to a vertex coloring
problem, which is a well-studied problem in graph theory. Although we use
a specific version of the greedy coloring algorithm it is possible that other coloring
algorithms may be used. We find that this approach leads to the most efficient solution
of the optimal box covering problem, but we also present two other methods based on breadth-first search
which address certain disadvantages of the first method, such as disconnected or
non-compact boxes.

We also compare our results with a number of methods introduced by others for studying this problem.
For example, Kim et al.~\cite{Kim} have used a variation of the random burning method,
where a random node serves as the seed of a box and neighboring unburned
nodes are assigned to this box. A similar method was applied for
edge-covering (instead of node-covering) which yields similar results
\cite{Sornette}.

\section{The greedy coloring algorithm}

We begin by recalling the original definition of box covering by Hausdorff \cite{Peitgen,Feder,HB}. For
a given network $G$ and box size $\ell_B$, a box is a set of
nodes where all distances $\ell_{ij}$ between any two nodes i and j in the box
are smaller than $\ell_B$. The minimum number of boxes required to cover
the entire network $G$ is denoted by $N_B$. 
For $\ell_B = 1$, $N_B$ is obviously equal to the size of the
network $N$, while $N_B=1$ for $\ell_B \ge
\ell_B^\mathrm{max}$, where $\ell_B^\mathrm{max}$ is the diameter
of the network (i.e. the maximum distance in the network) plus one.

\begin{figure}
\centerline{
 {\resizebox{6cm}{!} { \includegraphics{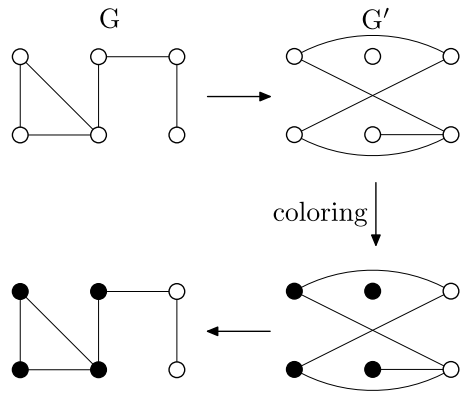}}}
} \caption {Illustration of the solution for the network covering problem
via mapping to the graph coloring problem. Starting
from $G$ (upper left panel) we construct the dual network $G'$ (upper right panel) for a given box size (here $\ell_B=3$), where two nodes are connected if they are
at a distance $\ell\geq\ell_B$. We use a greedy algorithm for vertex coloring in $G'$,
which is then used to determine the box covering in $G$, as shown in the plot.}
\label{aux}
\end{figure}

The ultimate goal of all box-covering algorithms is to locate the optimum
solution, i.e., to identify the minimum $N_B(\ell_B)$ value for
any given box size $\ell_B$.
We first demonstrate that this problem can be mapped to the graph
coloring problem, which is known to belong to the family of NP-hard
problems \cite{Garey}. This means
that an algorithm that can provide an exact solution
in a relatively short amount of time does not exist. This concept, though, enables us to treat
the box covering problem using known optimization approximations.
In order to find an approximation for the optimal solution for an arbitrary value of
$\ell_B$ we first construct a dual network $G'$, in which
two nodes are connected if the chemical distance between
them in $G$ (the original network) is greater or equal than $\ell_B$.
In Fig.~\ref{aux} we demonstrate an example of a network $G$
which yields such a dual network $G'$ for $\ell_B=3$ (upper row
of the figure).

Vertex coloring is a well-known procedure, where labels (or colors) are assigned to
each vertex of a network, so that no edge connects two
identically colored vertices. It is clear that such a coloring in
$G'$ gives rise to a natural box covering in the original
network $G$, in the sense that vertices of the same color will necessarily form a
box since the distance between them must be less than $\ell_B$.
Accordingly, the minimum number of boxes $N_B(G)$ is equal to the
minimum required number of colors (or the chromatic number) in the dual network $G'$,
$\chi(G')$, which is a famous problem in traditional graph theory.

In simpler terms, (a) if the distance between two nodes in $G$ is greater than $\ell_B$
these two neighbors cannot belong in the same box. According to the construction
of $G'$, these two nodes will be connected in $G'$ and thus they cannot have
the same color. Since they have a different color they will not belong in the
same box in $G$, which is our initial assumption. (b) On the contrary,
if the distance between two nodes in $G$ is less than $\ell_B$
it is possible that these nodes belong in the same box. In $G'$ these
two nodes will not be connected and it is allowed for these two nodes to carry the same color,
i.e. they may belong to the same box in $G$, (whether these nodes will actually be connected
depends on the exact implementation of the coloring algorithm, to be discussed later).

\begin{figure}
\centerline{
   {\resizebox{10cm}{!} { \includegraphics{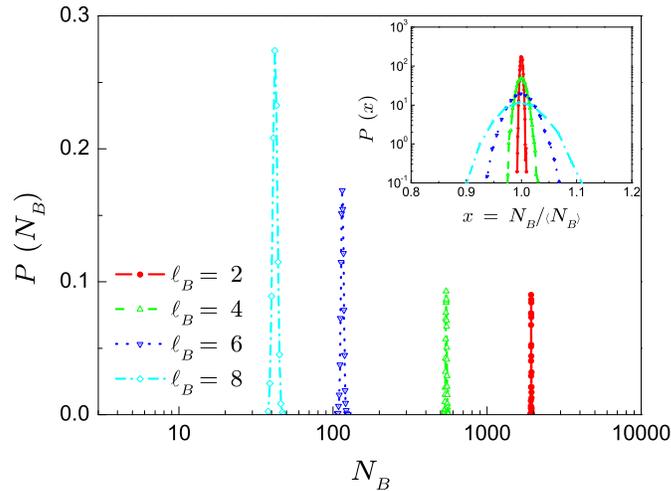}}}
} \caption {(color online) Probability distribution function $P(N_B)$
of the number of boxes $N_B$ for the greedy algorithm, applied to the
cellular network of {\it E.coli}. Different box sizes
$\ell_B$ are used as indicated in the plot.
Inset: PDFs of the normalized
quantity $N_B/\langle N_B\rangle$ in a semi-log plot for the greedy
algorithm, suggesting that
$P(N_B)$ follows a Gaussian distribution. } \label{PN}
\end{figure}

The exact solution for vertex coloring can only be achieved on small-size networks,
since the optimal number of colors in an arbitrary
graph is an NP-hard problem, as mentioned above,
and in general should be solved by
a brute-force approach \cite{Christofides,Wilf}. In practice, a greedy algorithm is widely adopted
to obtain an approximate solution \cite{Cormen} and this also works very well
for our case of box covering. 
We implement a simple version of the greedy algorithm
as follows: 1) Rank the nodes in a sequence, 2) Mark
each node with a free color, which is different from the colors of
its nearest neighbors in $G'$. 
The algorithm that follows both constructs the dual network $G'$ and assigns
the proper node colors for all $\ell_B$ values in one pass.
For this implementation we need a two-dimensional matrix $c_{i\ell}$ of size $N\times \ell_B^{\rm max}$,
whose values represent the color of node $i$ for a given box size $\ell=\ell_B$.
\begin{enumerate}
  \item Assign a unique id from 1 to N to all network nodes, without assigning any colors yet.
  \item For all $\ell_B$ values, assign a color value 0 to the node with id=1, i.e. $c_{1\ell}=0$.
  \item Set the id value $i=2$. Repeat the following until $i=N$.
  \begin{enumerate}
    \item Calculate the distance $\ell_{ij}$  from $i$ to all the
    nodes in the network with id $j$ less than $i$.
    \item Set $\ell_B=1$
    \item Select one of the unused colors $c_{j\ell_{ij}}$ from all nodes
    $j<i$ for which $\ell_{ij}\geq\ell_B$. This is the color $c_{i\ell_B}$ of node $i$ for the
    given $\ell_B$ value.
    \item Increase $\ell_B$ by one and repeat (c) until $\ell_B=\ell_B^\mathrm{max}$.
    \item Increase i by 1.
  \end{enumerate}
\end{enumerate}
This greedy algorithm is very efficient, since we can cover the network
with a sequence of box sizes $\ell_B$ performing only one network pass.

The results of the greedy algorithm may depend on the original coloring
sequence. In order to investigate the quality of the algorithm, we
randomly reshuffle the coloring sequence and apply the greedy
algorithm for 10,000 times on several different models and
real-world networks. In Fig.~\ref{PN} we present a typical example
for the PDFs of $N_B$ for the cellular network of {\it E.coli}.
The curves for all box sizes $\ell_B$ are narrow Gaussian distributions,
indicating that almost any implementation of the algorithm yields a
solution close to the optimal.

\begin{figure}
\centerline{
   {\resizebox{10cm}{!} { \includegraphics{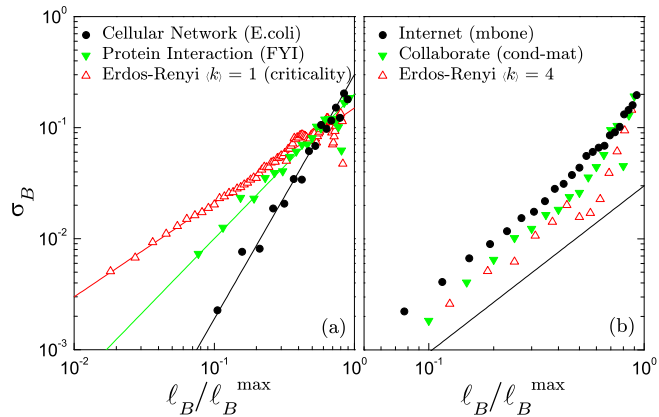}}}
} \caption {Normalized variance $\sigma_B$ of the greedy algorithm
for different box sizes $\ell_B$, for (a) fractal and (b)
non-fractal networks, where the box size $\ell_B$ is normalized by
the maximum box size $\ell_B^{\mathrm{max}}$. The slopes for
the fractal networks are (left to right): $\delta=$ 0.85, 1.3, 2.2.
For non-fractal networks: $\delta=1.5$.} \label{err}
\end{figure}

The uncertainty of the algorithm can be quantified via the
normalized variances $\sigma_B\equiv\langle\Delta
N_B^2\rangle^{1/2}/ \langle N_B\rangle$ of the PDFs. In Fig. \ref{err}, we present the
$\sigma_B$ dependence on the box size $\ell_B$ for both fractal (left
panel) and non-fractal (right panel) networks. Surprisingly, when $\ell_B<<\ell_{\rm max}$
all the
networks seem to exhibit a power-law dependence
\begin{equation}
\sigma_B\sim\ell_B^\delta \,,
\label{EQdelta}
\end{equation}
even for the case of non-fractal networks. In fractal networks the value
of $\delta$ depends on the network structure, while for non-fractal networks
$\delta$ seems to be constant with a value close to $1.5$.

\begin{figure}
\centerline{
   {\resizebox{6cm}{!} { \includegraphics{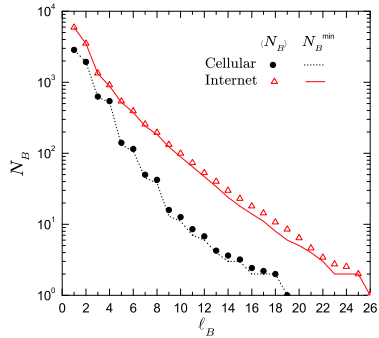}}}
} \caption {Comparison of the minimum $N_B^{\rm min}$ (line) and mean $\langle N_B \rangle$
(symbols) number of boxes for the greedy algorithm after 10,000
random reshuffles in real-world networks.
}
\label{comp0}
\end{figure}

Strictly speaking, the calculation of the fractal dimension $d_B$ through
the relation $N_B\sim \ell_B^{-d_B}$ is valid only for the minimum possible
value of $N_B$, for any given $\ell_B$ value, so an algorithm should aim to
find this minimum $N_B$.
For the greedy coloring algorithm it has been shown \cite{Cormen}
that it can identify a coloring sequence which
yields the optimal solution, i.e. the minimal value from the greedy algorithm
coincides with the optimal value. Obviously, there is no rule as to
when this minimum value has been actually reached. Yet, it is
still meaningful to compare the mean value $\langle N_B\rangle$
with the minimum value $N_B^{min}$ for our sample of 10,000 different
realizations. We present such a comparison for the cellular network (fractal)
and the Internet (non-fractal) in Fig.~\ref{comp0}. For all
$\ell_B$ values the difference between $\langle N_B\rangle$ and $N_B^{min}$ is very small
and the two values are almost indistiguishable from each other.
This result is significant for implementation purposes, by pointing out that
any realization of the above algorithm practically yields a quite accurate outcome.

The presented greedy algorithm is one of the simplest algorithms capable to
solve the exact coloring problem. The coloring problem is very important
in many fields, though, and consequently there is an enormous amount of studies
on this subject. In principle, any one of the suggested algorithmic solutions in the
literature can also be adopted for dealing with the box covering problem.

The form of the algorithm that was described above is the one that was used in
Refs.~\cite{shm, shm2} for the calculation of $N_B$ vs $\ell_B$.


\section{Burning algorithms}

The presented greedy-coloring algorithm provides at the same time high efficiency and
significant accuracy. A simpler approach, though, is to use more traditional
breadth-first algorithms. In the following sections we describe the basic simple burning
algorithm and introduce two alternative (more sophisticated) methods based on similar ideas.
We then proceed to compare these algorithms to the greedy-coloring algorithm.

\begin{figure}
\centerline{
{\resizebox{6cm}{!} { \includegraphics{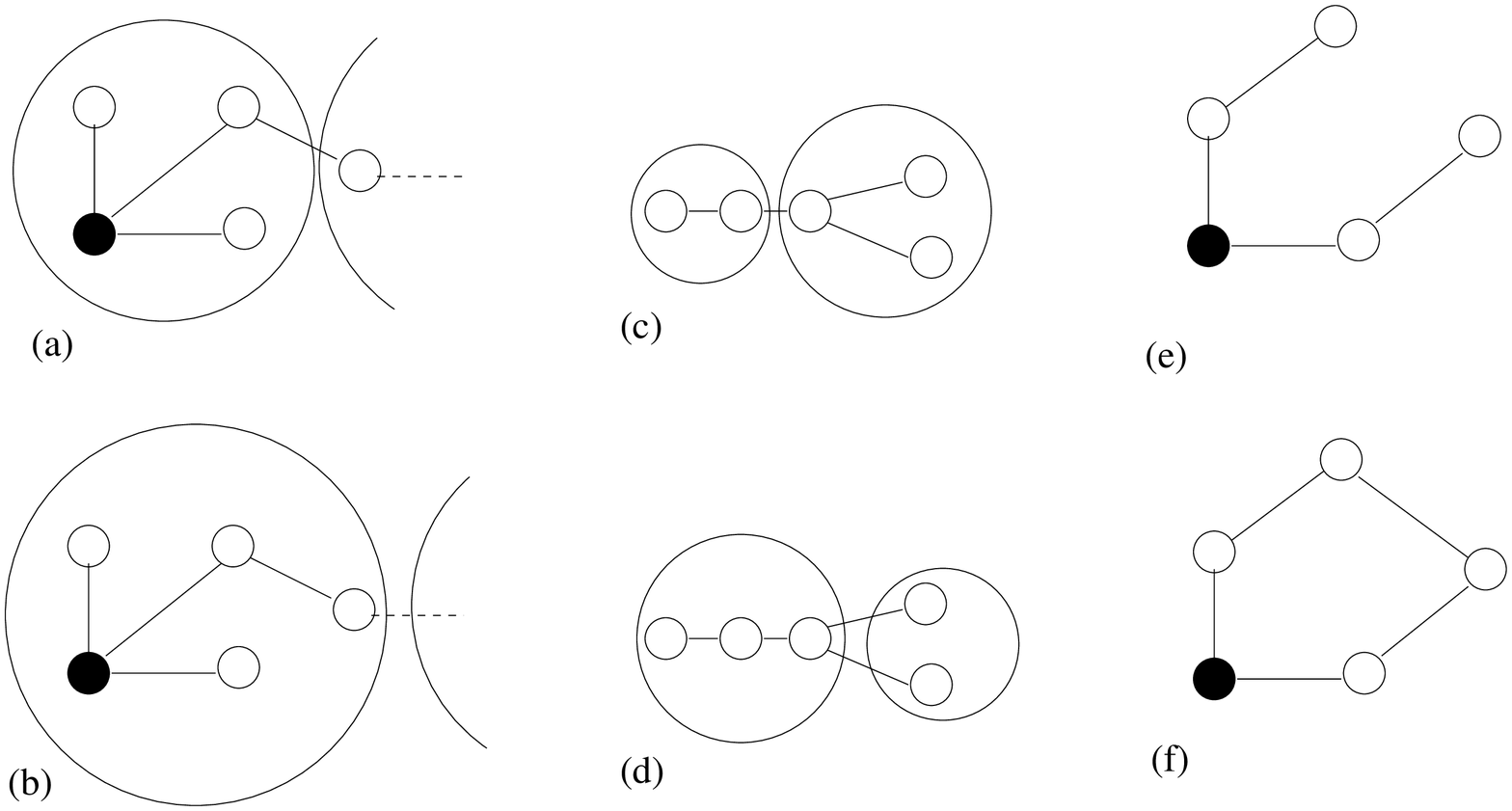}}}
} \caption {Our definitions for a box that is (a) non-compact for $\ell_B=3$, i.e. could include more nodes,
(b) compact, (c) connected, and (d) disconnected (the nodes in the right
box are not connected in the box). (e) For this box, the values $\ell_B=5$ and $r_B=2$
verify the relation $\ell_B=2r_B+1$. (f) One of the pathological cases
where this relation is not valid, since $\ell_B=3$ and $r_B=2$.} \label{definitions}
\end{figure}

In the following, we define a box to be `compact' when it includes the maximum
possible number of nodes, i.e. when there do not exist any other
network nodes that could be included in this box. A `connected' box means that
any node in the box can be reached from any other node in this box,
without having to leave this box. Equivalently, a `disconnected' box
denotes a box where certain nodes can be reached by other nodes in the box
only by visiting nodes outside this box. For a demonstration of these definitions
see Fig.~\ref{definitions}.

A short note on the definition of the distances used. A box of size $\ell_B$, according to our definition,
includes nodes where the distance between any pair of nodes is less than $\ell_B$. It is possible, though,
to grow a box from a given central node, so that all nodes in the box are within distance
less than a given box radius $r_B$ (the maximum distance from a central node).
For the original definition of the box, $\ell_B$ corresponds to the box diameter
(maximum distance between any two nodes in the box) plus one.
Thus, these two measures are related 
through $\ell_B = 2 r_B+1$. In general this relation is valid for random
configurations, but there may
exist specific cases, such as e.g. nodes in a cycle, where this 
equation is not exact (Fig.~\ref{definitions}).

\subsection{Burning with the diameter $\ell_B$, and the Compact-Box-Burning (CBB) algorithm}

A traditional geometrical approach is the so-called `burning'
algorithm (breadth-first search). The basic idea is to generate
a box by growing it from one randomly selected node towards its neighborhood
until the box is compact, or equivalently
that each box should include the maximum possible number of nodes.
The algorithm is quite simple and can be
summarized as follows:

\begin{enumerate}

\item{Choose a random uncovered node as the seed for a new box.}

\item{All uncovered nodes connected to the current box are tested for being
within distance $\ell_B$ from all the nodes currently in the box. Nodes that obey this
criterion are included in the box.}

\item{Repeat (ii) until there are no more nodes that can be added in this box.}

\item{Repeat (i)-(iii) until all nodes are covered.}

\end{enumerate}

Although this algorithm is quite easy to implement, it requires a very
long computational time. For this reason, we introduce a method that
yields the exact same results as the above algorithm, but is computationally
less intensive and can be executed much faster. We call this algorithm
Compact-Box-Burning or CBB.

The method can be better understood in geometrical terms (Fig.~\ref{FIGgeom_CBB}). 
We start from a random point and draw a circle with radius $\ell_B$. We then select a
random point within this circle and draw a circle with radius $\ell_B$ using this new center.
The union of the two circles includes all possible points that will eventually form the box.
Iteratively adding points from the union of all previous circles and drawing new circles
we eventually create a box where all the included points are within distance $\ell_B$ from
each other. For the case of a complex network,
we apply the following algorithm (see Fig.~\ref{FIGdraw_CBB}):

\begin{figure}
\centerline{
   {\resizebox{9cm}{!} { \includegraphics{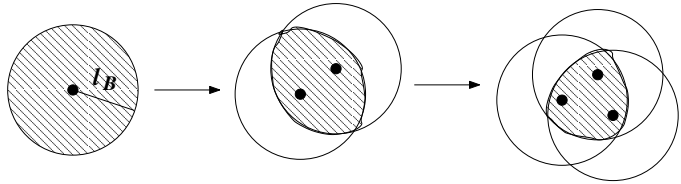}}}
} \caption { Two-dimensional geometrical analogue of the CBB algorithm. Initially we choose a random
point and consider the circle with radius $\ell_B$. We then choose another random point within
this circle which serves as a new circle center and calculate the union of these two circles. We continue by iteratively selecting
random centers for circles in the union of all the previous circles. 
} \label{FIGgeom_CBB}
\end{figure}

\begin{figure}
\centerline{
   {\resizebox{9cm}{!} { \includegraphics{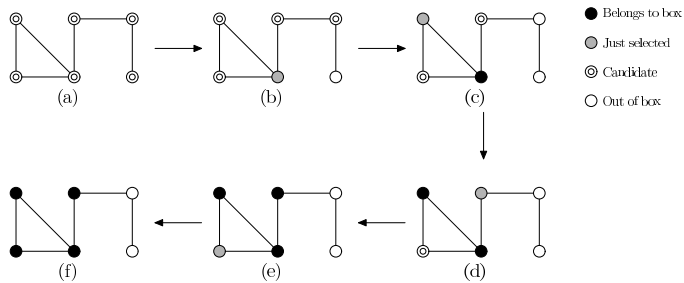}}}
} \caption {Illustration of the CBB algorithm for $\ell_B=3$. (a) Initially, all nodes are candidates for the box.
(b) A random node is chosen, and nodes at a distance further than $\ell_B$ from this
node are no longer candidates. (c) The node chosen in (b) becomes part of the box and another candidate
node is chosen. The above process is then repeated until the box is complete.} \label{FIGdraw_CBB}
\end{figure}

  \begin{enumerate}
   \item Construct the set $C$ of all yet uncovered nodes.

   \item Choose a random node $p$ from the candidate set $C$ and
   remove it from $C$.

   \item Remove from $C$ all nodes $i$ whose distance from $p$ is $\ell_{pi}\geq\ell_B$,
   since by definition they will not belong in the same box.

   \item Repeat steps (ii) and (iii) until the candidate set is empty.
  \end{enumerate}

The set of the chosen nodes ${p}$ forms a compact box. We then repeat the
above procedure until the entire network is covered.

\begin{figure}
\centerline{
   {\resizebox{10cm}{!} { \includegraphics{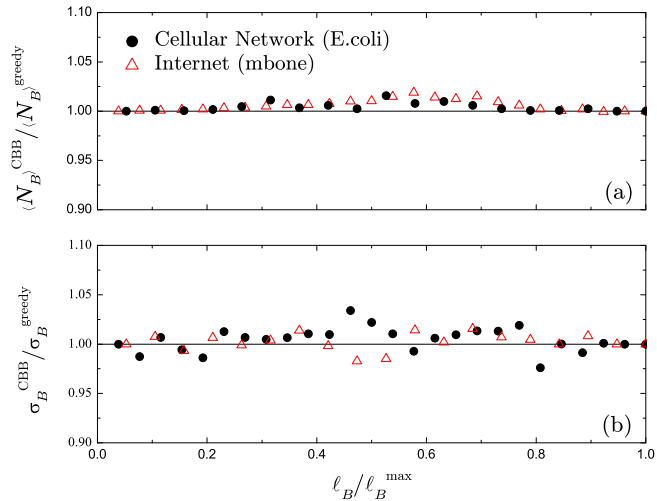}}}
} \caption {Comparison of (a) the mean number of boxes, $\langle N_B \rangle$, and
(b) the normalized variance, $\sigma_B$, between the greedy algorithm and
CBB. } \label{comp1}
\end{figure}

We also performed 10,000 realizations for the CBB algorithm and calculated
the mean value $\langle N_B\rangle$ and the normalized variance $\sigma_B$.
In Fig.~\ref{comp1} we compare the greedy algorithm with CBB for both fractal and
non-fractal networks. The value of $\langle N_B\rangle$ is roughly the same for both
algorithms, with the value from CBB slightly larger
(at most 2\%) than the one from the greedy algorithm. More interestingly, the
normalized variances are very close for these two
algorithms. This suggests that CBB provides results comparable with the greedy algorithm,
but CBB may be a bit simpler to implement.

\subsection{Burning with the radius $r_B$, and the Maximum-Excluded-Mass-Burning (MEMB) algorithm}

\begin{figure}
\centerline{
   {\resizebox{6cm}{!} { \includegraphics{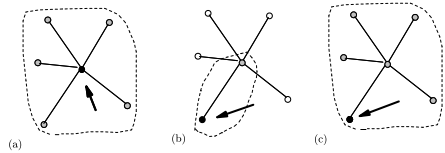}}}
} \caption {Burning with the radius $r_B$ from (a) a hub node or (b) a non-hub node results
in very different network coverage. In (a) we need just one box of $r_B=1$
while in (b) 5 boxes are needed to cover the same network. This is
an intrinsic problem when burning with the radius. (c) Burning
with the maximum distance $\ell_B$ (in this case $\ell_B=2r_B+1=3$) we avoid
this situation, since independently of the starting point we would
still obtain $N_B=1$.} \label{burning}
\end{figure}

The formal definition of boxes includes the maximum separation $\ell_B$ between any two nodes in
a box. However, it is possible to recover the same fractal properties of a network,
where a box can be defined as nodes within a radius $r_B$ from a central node.
Using this box definition and random central nodes, this burning algorithm yields the
optimal solution
for non scale-free homogeneous networks, since the choice of the central node is not important.
However, in inhomogeneous networks with wide-tailed degree distribution,
such as the scale-free networks,
this algorithm fails to achieve an optimal solution because of the hubs existence.
For example, Fig.~\ref{burning} demonstrates that burning with the radius
from non-hubs is much worse than burning from hubs. In scale-free networks,
when selecting a random node there is a high probability that this node will not be
a hub, but a low-degree node instead, which leads the network tiling far from the
optimal case. Additionally, a box
burning originating from a non-hub node is not compact, in the sense that
this box could contribute to a more efficient covering by incorporating
more uncovered nodes without violating the maximum distance criterion.
A variation of this algorithm for complex networks was presented in Ref.~\cite{Kim}.
In general, this method cannot directly provide the optimum coverage,
but it was shown that it finally yields the same fractal exponent $d_B$
as the greedy coloring algorithm. Since the most important feature of
similar studies is usually the calculation of the $d_B$ exponent this algorithm
can be very useful and, moreover, it is by far the easiest to implement.

\begin{figure}
\centerline{
   {\resizebox{9cm}{!} { \includegraphics{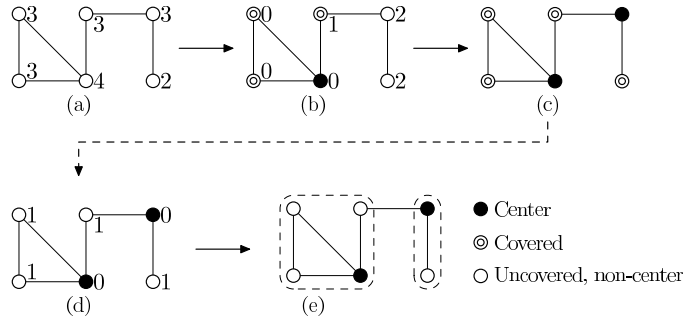}}}
} \caption {Illustration of the MEMB algorithm for $r_B=1$. {\it Upper row: Calculation of the box centers}
(a) We calculate the excluded mass for each node. (b) The node with maximum mass becomes a center
and the excluded masses are recalculated. (c) A new center is chosen. Now, the entire network is covered
with these two centers. {\it Bottom row: Calculation of the boxes} (d) Each box includes initially only the center.
Starting from the centers we calculate the distance of each network node to the closest center.
(e) We assign each node to its nearest box.} \label{FIGdraw_MEMB}
\end{figure}

To improve this completely random approach, we suggest an alternate strategy that
attempts to locate some optimal `central' nodes which will act as the burning origins for the boxes.
In principle one could use the hubs as box centers.
However, depending on the nature of the network, choosing the hubs may
not lead to the optimal solution because the hubs may be directly connected to
each other or share a large number of common nodes, and this choice
practically prohibits any low-degree
node to be a box center which in some cases may be beneficial. Burning from the
hubs represents a special case of the method that we will present, and
it may emerge naturally from this algorithm if this is indeed the
optimal way to cover the network. This is the case when hubs are not
directly connected.
In the following algorithm we
use the basic idea of box optimization, where we require that each box
should cover the maximum possible number of nodes.
For a given burning radius $r_B$, we define the ``excluded mass''
of a node as the number of uncovered nodes within a
chemical distance less than $r_B$. First, we calculate the excluded
mass for all the uncovered nodes. Then we seek to cover
the network with boxes of maximum excluded mass. The details of this
algorithm, which we call Maximum-Excluded-Mass-Burning or MEMB, are as follows
(see Fig.~\ref{FIGdraw_MEMB}):

\begin{enumerate}
  \item Initially, all the nodes are marked as uncovered and non-centers.
  \item For all non-center nodes (including the already covered nodes)
  calculate the excluded mass, and select the node $p$ with
  the maximum excluded mass as the next center.
  \item Mark all the nodes with chemical distance less than $r_B$ from $p$ as covered.
  \item Repeat steps (ii) and (iii) until all nodes are either covered or centers.
\end{enumerate}

Notice that the excluded mass
has to be updated in each step because it is possible that it has been modified during this step.
A box center can also be an already covered node, since it may lead to a largest box mass.
After the above procedure, the number of selected centers coincides with the number
of boxes $N_B$ that completely cover the network.
However, the non-center nodes have not yet been assigned to a given box.
This is performed in the next step:

\begin{enumerate}
  \item Give a unique box id to every center node.
  \item For all nodes calculate the ``central distance", which is the chemical distance to its
  nearest center. The central distance has to be less than $r_B$, and the center
  identification algorithm above guarantees that there will always exist such a center.
  Obviously, all center nodes have a central distance equal to 0.
  \item Sort the non-center nodes in a list according to increasing central distance.
  \item For each non-center node $i$, at least one of its neighbors has a
  central distance less than its own. Assign to $i$ the same id with this
  neighbor. If there exist several such neighbors, randomly select an id
  from these neighbors. Remove $i$ from the list.
  \item Repeat step (iv) according to the sequence from the list in step (iii) for all non-center nodes.
\end{enumerate}

\begin{figure}
\centerline{
   {\resizebox{10cm}{!} { \includegraphics{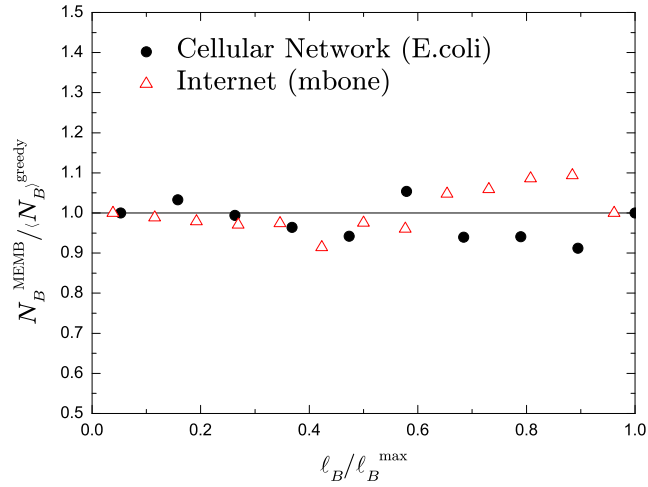}}}
} \caption {Comparison between the number of boxes obtained using MEMB, $N_B^{\mathrm{MEMB}}$,
and the mean number of boxes, $\langle N_B \rangle^\mathrm{greedy}$, obtained from the
greedy algorithm. } \label{comp2}
\end{figure}

For both the greedy coloring and the CBB algorithm the connectivity of boxes is
not guaranteed. That is, for some boxes there may not exist a path
inside the box that connects two nodes belonging in this box. The reason
is that some boxes may already include certain nodes that are crucial for the
optimization of other boxes. The MEMB algorithm, though, always yields
connected boxes and this is the most appropriate method when this condition
is required.

The MEMB algorithm is nearly deterministic, especially in the calculation of
the $N_B$ value. Randomness only enters in the order of choosing two nodes at equal
distance from two centers.
In order to directly compare the results with the greedy algorithm,
we convert the radius $r_B$ to the box-size $\ell_B$, according to $\ell_B = 2r_B+1$.
Fig.~\ref{comp2} shows that the calculated number of boxes using MEMB,
$N_B^{\mathrm{MEMB}}$, is also very similar to the mean value obtained from the
greedy algorithm, $\langle N_B\rangle^{\mathrm{greedy}}$.

The MEMB algorithm was used in Figs.~2 and 3 of Ref.~\cite{shm2} for the calculation of
hub-hub correlations, because in this case we want to isolate hubs in different boxes,
a behavior similar to the model introduced in that paper (through the quantity $\cal{E}(\ell_B)$ defined in \cite{shm2}).
Also, we used this algorithm for studying the evolution of conserved proteins in
the yeast protein interaction network \cite{unpublished}.

\subsection{Comparison between the different algorithms}

\begin{figure}
\centerline{
   {\resizebox{8cm}{!} { \includegraphics{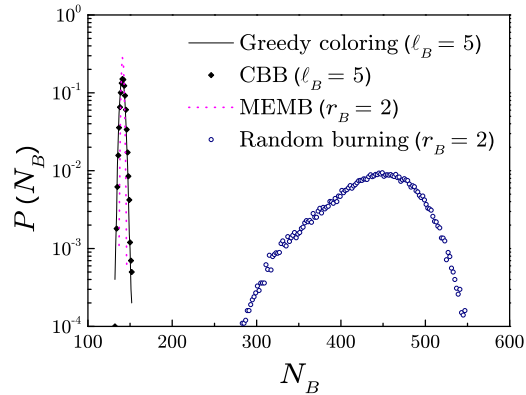}} }
} \caption {Comparison of the distribution of $N_B$ for $10^4$ realizations of the four network covering methods presented in
this paper. Notice that three of these methods yield very similar results with narrow distributions
and comparable minimum values, while the random burning algorithm fails to reach a value close to this minimum
(and yields a broad distribution).
 } \label{comp_all}
\end{figure}

A comparison between the greedy coloring, the CBB and MEMB algorithms with
the simple completely random burning with $r_B$ (Fig.~\ref{comp_all}) shows that the three methods,
except the random burning with $r_B$, are not sensitive to the specific realization used.
This is manifested in the very narrow distributions of $N_B$ and in the
minimum value of the distribution which is very similar in all three cases
(and very close to the average value, as well).
On the contrary, when we use the random burning algorithm with $r_B$ the corresponding distribution
is significantly wider and the mean value $\langle N_B \rangle$ is
much larger. Thus, a very large number of different
realizations is required for achieving the optimal coverage in this case.
Although the distributions in Fig.~\ref{comp_all} correspond to a given value
of $\ell_B$ (or equivalently $r_B$) the results are very similar for other $\ell_B$
values.

\begin{figure}
\centerline{
   {\resizebox{8cm}{!} { \includegraphics{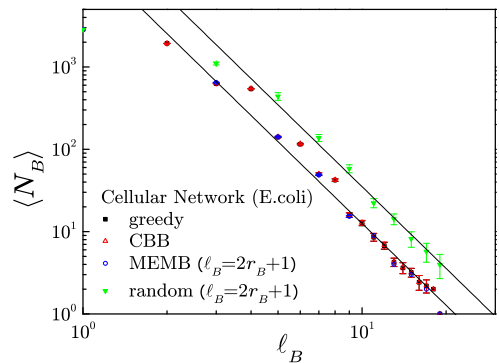}} }
} \caption {Comparison of the mean number of boxes $\langle N_B \rangle$ vs $\ell_B$
for the four presented algorithms. All methods yield the same value of the fractal dimension $d_B=3.5$.
 } \label{FIG_comp_db}
\end{figure}

Despite these differences, the calculation of the fractal dimension $d_B$ yields
the same value for all the presented algorithms (Fig.~\ref{FIG_comp_db}), indicating
that the scaling of the number of boxes is quite stable in all cases. Still,
for the random burning it is not clear how many different realizations are needed
in order for the average value to stabilize. Although from a practical point of
view burning with $r_B$ can still be used and give the correct dimension exponent $d_B$,
it is not clear whether the properties of the boxes will be the same as in the
optimal covering, e.g. whether applying renormalization to a network based on this covering
will be similar to the renormalized network obtained from the optimal tiling.

\section{Box-size correction}

In the usual box-covering techniques applied to regular fractals, as well as in
all the methods described above, the box-size $\ell_B$ denotes the maximum
possible distance within a box. Thus, it is always introduced as a cutoff
value, rather than a direct measurement. Although in homogeneous systems,
such as regular fractals, the difference may be indistinguishable, in many cases
concerning inhomogeneous networks the actual size of boxes can be
much smaller than this cutoff value $\ell_B$. This difference is
not expected to modify the asymptotic behavior of the scaling form
$N_B \sim \ell_B^{-d_B}$. However, measurement of the fractal dimension
$d_B$ in real-world networks usually requires faster convergence,
due to the small-world nature of many of them. Thus,
we introduce an alternative definition for the box size
$\ell_B^*$. This parameter corresponds now to the actual box size (after
we perform the network coverage in the usual way), and is defined
as the maximum distance inside the particular box plus one, which is of course always smaller
or equal to $\ell_B$. The average box size $\ell_B^*$ over all boxes is
used as a replacement of the previous cut-off size $\ell_B$,
and we replot the number of boxes $N_B(\ell_B^*)$ (whose maximum
diameter is still $\ell_B$) versus the average diameter $\ell_B^*$.
However, in order to obtain the correct box size
and be consistent with the $\ell_B^*$ definition,
the boxes have to be connected. Thus, we measure
$N_B(\ell_B^*)$ via the MEMB algorithm, as described above.

\begin{figure}
\centerline{
   {\resizebox{10cm}{!} { \includegraphics{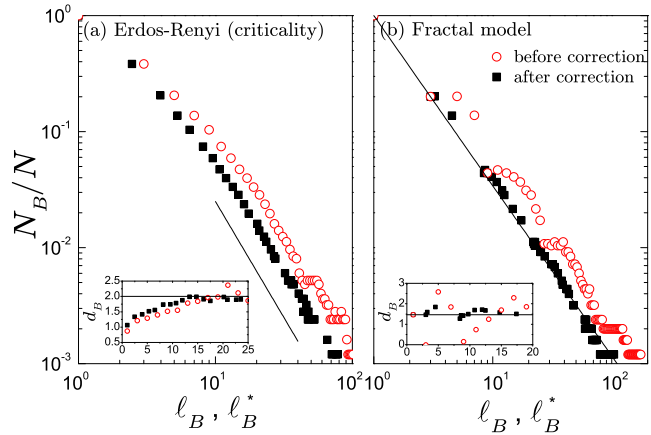}}}
} \caption {Comparison of the fractal dimension
before and after applying the box-size correction in (a) an Erdos-Renyi model at
criticality and (b) a fractal network model. The straight lines
correspond to the analytical predictions. Insets: Improvement of the fractal
dimension $d_B$ calculation as we increase the number of boxes used for this calculation.} \label{dimc}
\end{figure}

We test the improvement of this modification by applying the measurement
of $\ell_B^*$ to a couple of known examples.
The fractal dimension of Erdos-Renyi networks at criticality ($\langle k\rangle =1$) is known to be
$d_B=2$ (see e.g. \cite{Braunstein}). In Fig.~\ref{dimc}a we 
compare the numerical results before and after the size correction in such a network.
The measurement of fractality after the correction seems to converge faster to the analytical
prediction than the previous measurement. The improvement can be assessed by the inset
plot where the use of $\ell_B^*$ is shown that the theoretically
predicted value is achieved at smaller $\ell_B^*$. Furthermore, the proposed correction has smoothened the tail in
the plot, which may be crucial for the accurate determination of $d_B$, especially
for the small box sizes considered in real networks.

The improvement achieved is more prominent in the case of the fractal network model
proposed in \cite{shm2}.
Due to the construction process of this model, this network is highly modular
with very inhomogeneous distribution of the links in the modules. As a result,
the number of boxes for a given size $\ell_B$
fluctuates significantly and, as shown in Fig.~\ref{dimc}b, it is very difficult to
extract a reliable slope from the data. This discrete character has also been pointed
in Ref.~\cite{Sornette} where it is interpreted in terms of log-periodic oscillations
in $N_B$. The use of $\ell_B^*$, though,
leads to a very robust slope which is exhibited over almost the entire range.
As can be seen in the inset, $d_B$ is practically always equal to its theoretical value
when using the corrected value, in contrast to the uncorrected calculation where the
value of $d_B$ is more difficult to estimate.

\section{Summary}

In conclusion, we have shown that the box-covering method is equivalent
to vertex coloring in arbitrary networks. Based on this result,
we proposed a greedy algorithm for box covering, which was found
to be very accurate. A detailed analysis of the method was
performed to estimate the uncertainty of the algorithm. We also introduced
two geometric algorithms and compared them with the greedy
algorithm. We find that all of them result in a similar optimal number of
boxes. Finally, we showed that an alternate definition of the box size
$\ell_B$ can lead to a more precise measurement of a network's fractal dimension.

\ack{The authors acknowledge financial support from NSF grants}

\end{document}